\documentclass[12pt]{article}

\author{Yu.~M.~Zinoviev
       \thanks{E-mail address: Yurii.Zinoviev@ihep.ru} \\
        {\it Institute for High Energy Physics} \\
        {\it Protvino, Moscow Region, 142280, Russia}}
\title{On spin 3 interacting with gravity}

\date{}

\begin{document}

\maketitle

\begin{abstract}
Recently Boulanger and Leclercq have constructed cubic four derivative
$3-3-2$ vertex for interaction of spin 3 and spin 2 particles. This
vertex is trivially invariant under the gauge transformations of
spin 2 field, so it seemed that it could be expressed in terms of
(linearized) Riemann tensor. And indeed in this paper we managed to
reproduce this vertex in the form $R \partial \Phi \partial \Phi$,
where $R$ --- linearized Riemann tensor and $\Phi$ --- completely
symmetric third rank tensor. Then we consider deformation of this
vertex to $(A)dS$ space and show that such deformation produce
"standard" gravitational interaction for spin 3 particles (in linear
approximation) in agreement with general construction of Fradkin and
Vasiliev. Then we turn to the massive case and show that the same
higher derivative terms allows one to extend gauge invariant
description of massive spin 3 particle from constant curvature spaces
to arbitrary gravitational backgrounds satisfying $R_{\mu\nu} = 0$.
\end{abstract}

\thispagestyle{empty}
\newpage
\setcounter{page}{1}

\section*{Introduction}

The problem of constructing consistent interactions for high spin
particles is an old but still unsolved one (see e.g. reviews in 
\cite{Vas04,Sor04,BCIV05}). In this, one of the classical and
important tasks is investigation of gravitational interactions for
such particles. It was known for a long time that it is not possible
to construct standard gravitational interaction for massless high spin
$s \ge 5/2$ particles in flat Minkowski space \cite{AD79,WF80,BBD85}
(see also recent discussion in \cite{Por08}). At the same time, it has
been shown \cite{FV87,FV87a} that this task indeed has a solution in
$(A)dS$ space with non-zero cosmological term. The reason is that
gauge invariance, that turn out to be broken when one replace ordinary
partial derivatives by the gravitational covariant ones, could be
restored with the introduction of higher derivative corrections
containing gauge invariant Riemann tensor. These corrections have
coefficients proportional to inverse powers of cosmological constant
so that such theories do not have naive flat limit. But it is
perfectly possible to have a limit when both cosmological term and
gravitational coupling constant simultaneously go to zero in such a
way that only interactions with highest number of derivatives survive.

It is natural to suggest that in any realistic high spin theory (like
in superstring) most of high spin particles must be massive and their
gauge symmetries spontaneously broken. So a very important,
interesting and up to now poorly investigated problem is to find a
mechanism of spontaneous gauge symmetry breaking which could deform
massless particles in $(A)dS$ space into massive one in flat Minkowski
space. But if such mechanism does exist, one can apply the same line
of reasoning about massless limit in resulting theory. Namely, it is
natural to suggest that there exists a limit when both mass and
gravitational coupling constant simultaneously go to zero so that only
some interactions containing Riemann tensor survive.

In both cases above, the crucial point is the existence of cubic
higher derivative spin $s-s-2$ vertex containing (linearized) Riemann
tensor and two massless spin $s$ particles in flat Minkowski space.
While it is always possible to construct higher derivative vertex out
of gauge invariant "field strengths" \cite{DS87}, what we need here is
a vertex (with less derivatives) with non-trivial deformation of gauge
transformations. For spin $s=3$ case an appropriate candidate has been
constructed recently in \cite{BL06}. Really, the Lagrangian for four
derivatives $3-3-2$ vertex given in this paper does not look to be the
one we need here, but it is trivially invariant under the spin 2 gauge
transformations so it can be expressed in terms of (linearized)
Riemann tensor. Thus in the next section we reproduce this vertex in
the desired form. Then we consider deformation of this vertex into
$(A)dS$ space and show that such deformation does produce a standard
gravitational interaction for spin 3 particle (in linear
approximation). 

After that we turn to the main question we are interested in: could
non-zero mass play the same role as non-zero cosmological constant
(for the massive spin 5/2 case see \cite{Met06a}). To
our opinion the most natural and convenient formalism for
investigation of spontaneous gauge symmetry breaking is gauge
invariant description of massive high spin particles, which nicely
works both in flat Minkowski space as well as $(A)dS$ ones
\cite{KZ97,Zin01,Met06}. Due to large number of fields involved (four
for massive spin $s=3$ particle) complete investigation of
gravitational interaction for massive spin 3 requires a lot of work
and we leave it to the future (example for much simpler spin $s=2$
case see \cite{Zin06}),  so in this paper as a first step we consider
massive spin 3 particle in arbitrary gravitational background
satisfying its free equations, i.e. $R_{\mu\nu} = 0$. In this, only
terms containing full Riemann tensor survive which greatly simplifies
all calculations. 

In Section 3 we start with gauge invariant description of massive spin
3 particle in flat Minkowski space. In such formalism the problem of
switching on a gravitational interaction looks exactly the same as for
massless particles. Namely, when one replace ordinary partial
derivatives by the covariant ones the gauge invariance of the
Lagrangian turns out to be broken, but this non-invariance contains
terms with curvature tensor only leaving us the possibility to restore
gauge invariance with the help of higher derivative corrections. And
in the Section 4 we show that such restoration is indeed possible (at
least in the linear approximation) in this the same  cubic four
derivative $3-3-2$ vertex plays the main role.

\section{Cubic vertex 3-3-2 with four derivatives}

As we have already mentioned, crucial point is the existence of some
cubic higher derivative vertex in flat Minkowski space containing
(linearized) curvature tensor for spin 2 field. Appropriate vertex has
been constructed recently in \cite{BL06}. The form of the Lagrangian
written in this paper does not seem to be the one we are looking for,
but its trivial invariance under the spin 2 gauge transformations
shows that it can be reduced to the required form. In this Section we
reconstruct this vertex by using brute force method, namely we start
with the sum of free Lagrangians for massless spin 3 and spin 2
particles and their usual gauge transformations and consider the most
general four derivative cubic vertex and appropriate corrections to
gauge transformations.

The usual free Lagrangian for symmetric third rank tensor
$\Phi_{\mu\nu\alpha}$ has  the form:
\begin{equation}
{\cal L}_0 = - \frac{1}{2} \partial^\mu \Phi^{\nu\alpha\beta}
\partial_\mu \Phi_{\nu\alpha\beta} + \frac{3}{2} (\partial
\Phi)^{\mu\nu} (\partial \Phi)_{\mu\nu} - 3 (\partial \Phi)^{\mu\nu}
\partial_\mu \tilde{\Phi}_\nu + 
\frac{3}{2} \partial^\mu \tilde{\Phi}^\nu \partial_\mu
\tilde{\Phi}_\nu + \frac{3}{4} (\partial \tilde{\Phi}) (\partial
\tilde{\Phi})
\end{equation}
where $(\partial \Phi)_{\mu\nu} = \partial^\alpha
\Phi_{\mu\nu\alpha}$, $\tilde{\Phi}_\mu = g^{\alpha\beta}
\Phi_{\alpha\beta\mu}$. This Lagrangian is invariant under the
following gauge transformations:
\begin{equation}
\delta \Phi_{\mu\nu\alpha} = \partial_\mu \xi_{\nu\alpha} +
\partial_\nu \xi_{\mu\alpha} + \partial_\alpha \xi_{\mu\nu}
\end{equation}
where parameter $\xi_{\mu\nu}$ symmetric $\xi_{\mu\nu} = \xi_{\nu\mu}$
and traceless $\xi_\mu{}^\mu = 0$. Analogously, the well known
Lagrangian for symmetric second rank tensor $h_{\mu\nu}$ looks as
follows:
\begin{equation}
{\cal L}_0 = \frac{1}{2} \partial^\mu h^{\alpha\beta} \partial_\mu
h_{\alpha\beta} - (\partial h)^\mu (\partial h)_\mu + (\partial h)^\mu
\partial_\mu h - \frac{1}{2} \partial^\mu h \partial_\mu h
\end{equation}
being invariant under the gauge transformations with vector
parameter:
\begin{equation}
\delta h_{\mu\nu} = \partial_\mu \xi_\nu + \partial_\nu \xi_\mu
\end{equation}

The most general four derivative cubic vertex which is trivially
invariant under the spin 2 gauge transformations can be written
(symbolically) as:
$$
{\cal L} \sim R \partial^2 \Phi \Phi \oplus R \partial \Phi \partial
\Phi
$$
where $R$ denotes linearized Riemann tensor:
\begin{equation}
R_{\mu\nu,\alpha\beta} = \partial_\mu \partial_\alpha h_{\nu\beta} -
\partial_\nu \partial_\alpha h_{\mu\beta} - \partial_\mu
\partial_\beta h_{\nu\alpha} + \partial_\nu \partial_\beta
h_{\mu\alpha}
\end{equation}
In this, the most general ansatz for the gauge transformation
corrections has the form:
$$
\delta \Phi \sim \partial R \xi \oplus R \partial \xi \qquad
\delta h \sim \partial^3 \Phi \xi \oplus \partial^2 \Phi \partial \xi
$$
Note, that we will not introduce corrections to the gauge
transformations containing higher derivatives on gauge transformation
parameter $\xi$  because such terms (in general) change the whole
structure of constraints which leads to a change in number of physical
degrees of freedom.

There are a number of ambiguities which arise in such straightforward
construction (see e.g. discussion in \cite{BFPT06}):
\begin{itemize}
\item First of all, the interacting Lagrangian and gauge
transformations are always defined up to possible field redefinitions,
which do not change physical content of the theory. In this particular
case, we have a number of field redefinitions of the form:
$$
\Phi \Rightarrow \Phi \oplus R \Phi
$$
\item Due to gauge invariance of free Lagrangians, gauge
transformations in this approximation are defined up total divergency:
$$
\delta \Phi \sim \partial (R \xi), \qquad
\delta h \sim \partial ( \partial^2 \Phi \xi)
$$
\item Working with higher derivative gauge transformations one
always faces a number of trivial symmetries, i.e. gauge
transformations leaving sum of free Lagrangians invariant which do not
correspond to any non-trivial interactions \cite{BBD86}.
\item At last, not all the terms in the interacting Lagrangians which
could be constructed are independent because as usual such Lagrangians
are defined up total divergency and  there are many groups of terms
which could be combined into total divergency.
\end{itemize}

We resolve these ambiguities in a threefold way. First of all, we try
to lower the number of derivatives in the equations of motions. In
general, working with four derivative Lagrangians we get equations
containing up to the four derivatives on the fields, but in this case
we
managed to bring resulting Lagrangian into the form when equations
contain at most third derivatives. Then we use remaining freedom to
make gauge transformations as simple as possible. At last, some
freedom that still remain we will use in the next Section to bring two
derivative terms into the form of "standard" gravitational
interaction.

Now we require that total Lagrangian be invariant under the corrected
gauge transformations up to the terms bilinear in fields and also that
algebra of gauge transformations be closed in the lowest order. The
final Lagrangian (a result of very lengthy calculations which we will
not reproduce here) has the form:
\begin{eqnarray}
48 M^3 {\cal L}_1 &=& 8 R_{\mu\nu,\alpha\beta} [ \partial^\mu
\Phi^{\nu\rho\sigma} \partial^\alpha \Phi^{\beta\rho\sigma} +
2 \partial^\mu \Phi^{\alpha\rho\sigma} \partial^\rho
\Phi^{\sigma\nu\beta} - 
 \partial^\rho \Phi^{\sigma\mu\alpha} \partial^\sigma
\Phi^{\rho\nu\beta} - \partial^\nu \Phi^{\mu\alpha\rho}
\partial^\beta \tilde{\Phi}^\rho - \nonumber \\
 && \qquad -  \partial^\nu \Phi^{\mu\beta\rho} \partial^\rho
\tilde{\Phi}^\alpha - \partial^\mu \tilde{\Phi}^\nu \partial^\alpha
\tilde{\Phi}^\beta + 
 (\partial \Phi)^{\mu\alpha} (\partial \Phi)^{\nu\beta} - 3
(\partial \Phi)^{\mu\alpha} \partial^\nu \tilde{\Phi}^\beta ] +
\nonumber \\
 && + 4 R_{\mu\nu} [ - \partial^\mu \Phi^{\alpha\beta\rho}
\partial^\nu \Phi^{\alpha\beta\rho} + 6 \partial^\mu
\Phi^{\alpha\beta\rho} \partial^\alpha \Phi^{\nu\beta\rho} - 2
\partial^\alpha \Phi^{\mu\beta\rho} \partial^\beta
\Phi^{\nu\alpha\rho} + \nonumber \\
 && \qquad + \partial^\alpha \Phi^{\mu\beta\rho}
\partial^\alpha \Phi^{\nu\beta\rho} -
4 \partial^\alpha \Phi^{\mu\nu\beta} (\partial
\Phi)^{\alpha\beta} - 6 \partial^\mu \Phi^{\nu\alpha\beta}
\partial^\alpha \tilde{\Phi}^\beta + 2 \partial^\alpha
\Phi^{\mu\nu\beta} \partial^\alpha \tilde{\Phi}^\beta + \nonumber \\
 && \qquad + 2 \partial^\mu \tilde{\Phi}^\alpha \partial^\nu
\tilde{\Phi}^\alpha
- 10 \partial^\mu \tilde{\Phi}^\alpha \partial^\alpha
\tilde{\Phi}^\nu + 4 \partial^\alpha \Phi^{\mu\nu\beta} \partial^\beta
\tilde{\Phi}^\alpha + 9 \partial^\mu \tilde{\Phi}^\nu (\partial
\tilde{\Phi}) - \nonumber \\
 && \qquad - \partial^\alpha \tilde{\Phi}^\mu \partial^\alpha
\tilde{\Phi}^\nu
- 2 (\partial \Phi)^{\mu\alpha} \partial^\nu
\tilde{\Phi}^\alpha + 2 (\partial \Phi)^{\mu\alpha} \partial^\alpha
\tilde{\Phi}^\nu ] + \nonumber \\
 && + R [ - 2 \partial^\mu \Phi^{\nu\alpha\beta} \partial^\mu
\Phi^{\nu\alpha\beta} - 2 \partial^\mu \Phi^{\nu\alpha\beta}
\partial^\nu \Phi^{\mu\alpha\beta} + 12 (\partial \Phi)^{\mu\nu}
\partial^\mu \tilde{\Phi}^\nu - \nonumber \\
 && \qquad - 4 \partial^\mu \tilde{\Phi}^\nu
\partial^\mu \tilde{\Phi}^\nu + 4 \partial^\mu \tilde{\Phi}^\nu
\partial^\nu \tilde{\Phi}^\mu - 9 (\partial \tilde{\Phi}) (\partial
\tilde{\Phi}) ] \label{Lag}
\end{eqnarray}
(let us stress once again that it is not in any way unique). Here,
using the fact that cubic four derivative vertex for bosonic fields
must have dimension $\frac{1}{m^3}$, we introduce characteristic scale
$M$. In this, appropriate corrections to gauge transformations look as
follows:
\begin{eqnarray}
6 M^3 \delta \Phi_{\mu\nu\alpha} &=& 2 R_{\mu\beta\nu\rho}
\partial_{[\alpha} \eta_{\beta]\rho} - g_{\mu\nu}
R_{\alpha\beta\rho\sigma} \partial^\sigma \eta^{\beta\rho} +
(\mu\nu\alpha) \\
3 M^3 \delta h_{\mu\nu} &=& [ \partial_\alpha \partial_\beta
\Phi_{\mu\nu\rho} - \partial_\alpha \partial_\mu \Phi_{\nu\rho\beta}
- \partial_\alpha \partial_\nu  \Phi_{\mu\rho\beta} ]
 \partial^{[\alpha} \eta^{\rho]\beta} 
\end{eqnarray}
Note that the $h$-transformations have exactly the same form as in
\cite{BL06}, while $\Phi$-ones differ by total divergency.
Thus, we indeed reproduce cubic vertex of this paper in the form very
well suitable for investigations of gravitational spin 3 interactions.

\section{Deformation to $(A)dS$ space}

In this Section we consider deformation of four derivative vertex
presented above into $(A)dS$ space.\footnote{As far as we know, for
the first time this task was considered by M. A. Vasiliev in the
middle of 80's (unpublished).} $(A)dS$ space is a constant curvature
space without torsion or non-metricity, so the main difference from
Minkowski space is that (covariant) derivatives do not commute any
more. Our conventions will be:
\begin{equation}
[ D_\mu, D_\nu ] v_\alpha = - \kappa (g_{\mu\alpha} v_\nu -
g_{\nu\alpha} v_\mu), \qquad \kappa = \frac{2 \Lambda}{(d-1)(d-2)}
\end{equation}
For simplicity, in this Section we restrict ourselves with $d=4$
space. Non-commutativity of derivatives leads to some ambiguity even
in the form of free Lagrangians and also requires addition to
Lagrangian some mass-like terms to keep gauge invariance intact. We
will use the following concrete form for the massless spin 3
Lagrangian in $(A)dS_4$ space:
\begin{eqnarray}
{\cal L}_0 &=& - \frac{1}{2} D^\mu \Phi^{\nu\alpha\beta}
D_\mu \Phi_{\nu\alpha\beta} + \frac{3}{2} (D \Phi)^{\mu\nu} (D
\Phi)_{\mu\nu} - 3 (D \Phi)^{\mu\nu} D_\mu \tilde{\Phi}_\nu + 
\frac{3}{2} D^\mu \tilde{\Phi}^\nu D_\mu \tilde{\Phi}_\nu + \nonumber
\\
 && + \frac{3}{4} (D \tilde{\Phi}) (D \tilde{\Phi})
- \frac{\kappa}{2} \Phi^{\mu\nu\alpha} \Phi_{\mu\nu\alpha} + 9
\kappa \tilde{\Phi}^\mu \tilde{\Phi}_\mu
\end{eqnarray}
as well as the following one for massless spin 2 particle:
\begin{eqnarray}
{\cal L}_0 &=& \frac{1}{2} D^\mu h^{\alpha\beta} D_\mu h_{\alpha\beta}
- \frac{1}{2} (D h)^\mu (D h)_\mu - \frac{1}{2} D^\mu h^{\alpha\beta}
D_\alpha h_{\mu\beta}  + (D h)^\mu D_\mu h - \frac{1}{2} D^\mu h D_\mu
h + \nonumber \\
 && + \kappa (h^{\mu\nu} h_{\mu\nu} - h^2)
\end{eqnarray}

Next, we will use the following generalization of linearized Riemann
tensor:
\begin{eqnarray}
2 R_{\mu\nu,\alpha\beta} &=& D_\mu D_\alpha h_{\nu\beta} - D_\nu
D_\alpha h_{\mu\beta} - D_\mu D_\beta h_{\nu\alpha} + D_\nu D_\beta
h_{\mu\alpha} + \nonumber \\
 && + D_\alpha D_\mu h_{\nu\beta} - D_\alpha D_\nu h_{\mu\beta} -
D_\beta D_\mu h_{\nu\alpha} + D_\beta D_\nu h_{\mu\alpha} - \nonumber
\\
 && - 2\kappa (g_{\mu\alpha} h_{\nu\beta} - g_{\nu\alpha} h_{\mu\beta}
- g_{\mu\beta} h_{\nu\alpha} + g_{\nu\beta} h_{\mu\alpha} )
\end{eqnarray}
Besides being gauge invariant under covariant gauge transformations
$\delta h_{\mu\nu} = D_\mu \xi_\nu + D_\nu \xi_\mu$, such definition
allows us to keep usual properties of Riemann tensor such as:
$$
R_{\mu\nu,\alpha\beta} = R_{\alpha\beta,\mu\nu}, \qquad
D^\mu R_{\mu\nu,\alpha\beta} = D_\alpha R_{\nu\beta} - D_\beta
R_{\nu\alpha}
$$
At last, there is an ambiguity in the generalization of
$h$-transformations containing second derivatives of $\Phi$, but there
is no any evident choice here. So we leave this transformation as is
and take this ambiguity into account introducing explicit corrections
of the form $\delta h \sim \Phi D \xi$ later on.

As usual, if we replace all derivatives in our four derivative vertex
and gauge transformations by the covariant ones, we loose gauge
invariance. This non-invariance appears due to non-zero commutator of
covariant derivatives and as a result contains lower derivative
terms. Explicitly it looks as follows (again this results from some
lengthy calculations we omit here):
\begin{eqnarray}
6 \frac{M^3}{\kappa} \delta {\cal L}_1 &=& D_\mu R_{\alpha\beta} ( 14
\Phi^{\alpha\beta\nu} \xi^{\mu\nu} - 28 \Phi^{\mu\alpha\nu}
\xi^{\beta\nu} + 20 \tilde{\Phi}^\mu \xi^{\alpha\beta} + 4
\tilde{\Phi}^\alpha \xi^{\mu\beta} ) - \nonumber \\
 &&  - 16 D_\rho R_{\mu\nu,\alpha\beta} \Phi^{\rho\mu\alpha}
\xi^{\nu\beta}  + D_\mu R ( 4 \Phi^{\mu\alpha\beta} \xi_{\alpha\beta}
- 11 \tilde{\Phi}_\nu \xi^{\mu\nu} ) + \nonumber \\
 && + R_{\mu\nu,\alpha\beta} ( 10 D^\nu \Phi^{\mu\alpha\rho}
\xi^{\beta\rho} - 40 (D \Phi)^{\mu\alpha} \xi^{\nu\beta} + 42 D^\mu
\tilde{\Phi}^\alpha \xi^{\nu\beta}) + \nonumber \\
 && + R_{\mu\nu} ( 28 D^\alpha \Phi^{\mu\nu\beta} \xi_{\alpha\beta}
- 24 D^\mu \Phi^{\nu\alpha\beta} \xi_{\alpha\beta} + 4
D_\alpha \tilde{\Phi}^\mu \xi^{\alpha\nu} + 4 D^\mu
\tilde{\Phi}_\alpha \xi^{\nu\alpha}  - \nonumber \\
 && - 24 (D \Phi)^{\mu\alpha} \xi^{\nu\alpha} + 11 (D \tilde{\Phi})
\xi^{\mu\nu} ) + R (18 (D \Phi)^{\mu\nu} \xi_{\mu\nu} - 22
D^\mu \tilde{\Phi}^\nu \xi_{\mu\nu} )
\end{eqnarray}

Now to compensate this non-invariance we have to introduce lower
derivative corrections to cubic vertex as well as to gauge 
transformations. From the last formula we see that "effective"
gravitational coupling constant turns out to be $\frac{\kappa}{M^3}
\sim \frac{1}{m_{pl}}$. In what follows we set this constant to be 1.
We proceed as follows. First of all, we introduce all
corrections which appears when one replace background metric by the
"dynamical" one:
\begin{equation}
g_{\mu\nu} \rightarrow g_{\mu\nu} + h_{\mu\nu}
\end{equation}
In-particular, this requires generalization of tracelessness condition
for parameter $\xi_{\mu\nu}$:
\begin{equation}
\xi_{\mu\mu} = 0 \rightarrow \xi_{\mu\mu} - h^{\mu\nu} \xi_{\mu\nu} =
0
\end{equation}
Then we replace all derivatives by the "fully" covariant ones, e.g.:
\begin{equation}
D_\mu \xi_{\nu\alpha} \rightarrow D_\mu \xi_{\nu\alpha} -
\Gamma_{\mu\nu}{}^\beta \xi_{\beta\alpha} - \Gamma_{\mu\alpha}{}^\beta
\xi_{\nu\beta}
\end{equation}
where we introduce linearized Christoffel symbols:
\begin{equation}
\Gamma_{\mu\nu,\alpha} = \frac{1}{2} ( D_\mu h_{\nu\alpha} + D_\nu
h_{\mu\alpha} - D_\alpha h_{\mu\nu})
\end{equation}
Also we introduce transformations for the $\Phi$ field corresponding
to standard general coordinate transformations for third rank tensor:
\begin{equation}
\delta \Phi_{\mu\nu\alpha} = \xi^\beta D_\beta \Phi_{\mu\nu\alpha} +
D_\mu \xi^\beta \Phi_{\beta\nu\alpha} + D_\nu \xi^\beta
\Phi_{\mu\beta\alpha} + D_\alpha \xi^\beta \Phi_{\mu\nu\beta}
\end{equation}
Besides these corrections reproducing (in linear approximation)
standard gravitational interaction for massless spin 3 particle we
need two more corrections to achieve full gauge invariance under both
$\xi_{\mu\nu}$ and $\xi_\mu$ transformations. At first, we have to add
non-minimal terms to the Lagrangian (again their explicit form is not
unique):
\begin{eqnarray}
\Delta {\cal L} &=& - 4 R_{\mu\nu,\alpha\beta} \Phi^{\mu\alpha\rho}
\Phi^{\nu\beta\rho} + \frac{1}{4} R_{\mu\nu} ( - 21
\Phi^{\mu\alpha\beta} \Phi^{\nu\alpha\beta} + 22 \Phi^{\mu\nu\alpha}
\tilde{\Phi}_\alpha + 11 \tilde{\Phi}^\mu \tilde{\Phi}^\nu) +
\nonumber \\
 && + \frac{1}{8} R ( 14 \Phi_{\mu\nu\alpha}{}^2 - 19
\tilde{\Phi}_\mu{}^2)
\end{eqnarray}
At last, we need some corrections to $h$-transformations we mentioned
earlier:
\begin{equation}
\delta h_{\mu\nu} = - \Phi_{\mu\alpha\beta} D_\nu \xi^{\alpha\beta} +
(\mu \leftrightarrow \nu) - \Phi_{\mu\nu\alpha} (D \xi)^\alpha
\end{equation}

\section{Gauge invariant description of massive spin 3}

In what follows we will use gauge invariant description of massive
spin 3 particles in Minkowski space \cite{KZ97}, though it is possible
to generalize these results to $(A)dS$ space as well \cite{Zin01}.
Besides third rank tensor $\Phi_{\mu\nu\alpha}$ we need three more
fields now: symmetric second rank tensor $f_{\mu\nu}$, vector $A_\mu$
and scalar $\varphi$. We start with the sum of massless Lagrangians
for all four fields (we use non-canonical normalization of kinetic
terms to simplify mass terms and gauge transformations):
\begin{eqnarray}
{\cal L}_{02} &=& - \frac{1}{2} \partial^\mu \Phi^{\nu\alpha\beta}
\partial_\mu \Phi_{\nu\alpha\beta} + \frac{3}{2} (\partial
\Phi)^{\mu\nu} (\partial \Phi)_{\mu\nu} - 3 (\partial \Phi)^{\mu\nu}
\partial_\mu \tilde{\Phi}_\nu + \frac{3}{2} \partial^\mu
\tilde{\Phi}^\nu \partial_\mu \tilde{\Phi}_\nu + \frac{3}{4} (\partial
\tilde{\Phi}) (\partial \tilde{\Phi}) + \nonumber \\
 && + 3(\frac{1}{2} \partial^\mu f^{\alpha\beta} \partial_\mu
f_{\alpha\beta} - (\partial f)^\mu (\partial f)_\mu + (\partial f)^\mu
\partial_\mu f - \frac{1}{2} \partial^\mu f \partial_\mu f) -
\nonumber \\
 && - \frac{6(d+1)}{d} ( \partial^\mu A^\nu \partial_\mu A_\mu -
\partial^\mu A^\nu \partial_\nu A_\mu) + \frac{18(d+1)}{d-2}
\partial^\mu \varphi \partial_\mu \varphi
\end{eqnarray}
here and in what follows $d \ge 4$ and their usual gauge
transformations:
\begin{equation}
\delta_{01} \Phi_{\mu\nu\alpha} = \partial_\mu \xi_{\nu\alpha} +
\partial_\nu \xi_{\mu\alpha} + \partial_\alpha \xi_{\mu\nu}, \qquad
\delta_{01} f_{\mu\nu} = \partial_\mu \xi_\nu + \partial_\nu \xi_\mu,
\qquad \delta_{01} A_\mu = \partial_\mu \xi
\end{equation}
To obtain gauge invariant description of massive spin 3 particle we
introduce cross terms with one derivative:
\begin{eqnarray}
\frac{1}{m} {\cal L}_{01} &=& - \frac{3}{2} [2 \Phi^{\mu\nu\alpha}
\partial_\mu f_{\nu\alpha} - 4 \tilde{\Phi}^\mu (\partial f)_\mu +
\tilde{\Phi}^\mu \partial_\mu f ] + \nonumber \\
 && + \frac{12(d+1)}{d} (f^{\mu\nu} \partial_\mu A_\nu - f
 (\partial A)) - \frac{36(d+1)^2}{(d-2)^2} A^\mu \partial_\mu \varphi
\end{eqnarray}
as well as appropriate mass terms into the Lagrangian:
\begin{eqnarray}
\frac{1}{m^2} {\cal L}_{00} &=& \frac{1}{2} \Phi_{\mu\nu\alpha}{}^2 -
\frac{3}{2} \tilde{\Phi}_\mu{}^2 + \frac{6(d+1)}{d} \tilde{\Phi}^\mu
A_\mu - \frac{6(d+1)(d+2)}{d^2} A_\mu{}^2 + \nonumber \\
 && + \frac{9}{4} f^2 - \frac{18(d+1)}{d-2} f \varphi +
\frac{36(d+1)^2}{(d-2)^2} \varphi^2
\end{eqnarray}
as well as appropriate corrections to gauge transformations:
\begin{eqnarray}
\delta_{00} \Phi_{\mu\nu\alpha} &=& \frac{2m}{d} (g_{\mu\nu}
\xi_\alpha + g_{\mu\alpha} \xi_\nu + g_{\nu\alpha} \xi_\mu), \qquad
\delta_{00} A_\mu = m \xi_\mu \nonumber \\
\delta_{00} f_{\mu\nu} &=& m \xi_{\mu\nu} + \frac{4(d+1)}{d(d-1)} m
g_{\mu\nu} \xi, \qquad \delta_{00} \varphi = m \xi
\end{eqnarray}

Let us recall a relation of such gauge invariant description with
usual one. Using gauge transformations with parameters $\xi_{\mu\nu}$,
$\xi_\mu$ and $\xi$ one can always choose a gauge where $f_{\mu\nu} -
\frac{1}{d} g_{\mu\nu} f = 0$, $A_\mu = 0$ and $\varphi = 0$. This
leaves us with the main field $\Phi_{\mu\nu\alpha}$ and one scalar
auxiliary field $f$ (the trace of $f_{\mu\nu}$) with the Lagrangian:
\begin{eqnarray}
{\cal L} &=& - \frac{1}{2} \partial^\mu \Phi^{\nu\alpha\beta}
\partial_\mu \Phi_{\nu\alpha\beta} + \frac{3}{2} (\partial
\Phi)^{\mu\nu} (\partial \Phi)_{\mu\nu} - 3 (\partial \Phi)^{\mu\nu}
\partial_\mu \tilde{\Phi}_\nu + \frac{3}{2} \partial^\mu
\tilde{\Phi}^\nu \partial_\mu \tilde{\Phi}_\nu + \frac{3}{4} (\partial
\tilde{\Phi}) (\partial \tilde{\Phi}) + \nonumber \\
 && - \frac{3(d-1)(d-2)}{d^2} \partial^\mu f \partial_\mu f -
\frac{3m(d-2)}{2d} \tilde{\Phi}^\mu \partial_\mu f + m^2 [\frac{1}{2}
\Phi_{\mu\nu\alpha}{}^2 - \frac{3}{2} \tilde{\Phi}_\mu{}^2 +
\frac{9}{4} f^2 ]
\end{eqnarray}

Now if we move to non-trivial gravitational background\footnote{For
massive spin 2 in gravitational background see
\cite{BGKP99,BKP99,BGP00}} satisfying
$R_{\mu\nu} = 0$, where
\begin{equation}
[ D_\mu, D_\nu ] v_\alpha = R_{\mu\nu,\alpha\beta} v^\beta
\end{equation}
we loose gauge invariance
\begin{equation}
\delta {\cal L}_0 = 3 R_{\mu\nu,\alpha\beta} ( 4 D^\nu
\Phi^{\mu\alpha\rho} \xi^{\beta\rho} - 2 (D \Phi)^{\mu\alpha}
\xi^{\nu\beta} + 4 D^\mu \tilde{\Phi}^\alpha \xi^{\nu\beta} 
- 2 f^{\mu\alpha} \xi^{\nu\beta} - 2 D^\nu f^{\mu\alpha} \xi^\beta )
\end{equation}
One could try to restore gauge invariance by adding non-minimal terms,
which in this simple case are just:
\begin{equation}
\Delta {\cal L}_0 = R_{\mu\nu,\alpha\beta} ( b_1 \Phi^{\mu\alpha\rho}
\Phi^{\nu\beta\rho} + b_2 f^{\mu\alpha} f^{\nu\beta} )
\end{equation}
but it is easy to check that it is impossible.

\section{Massive spin 3 in gravitational background}

In this Section we show that introducing higher derivative
corrections to Lagrangian and gauge transformations it is possible to
obtain gauge invariant description of massive spin 3 particle in
arbitrary gravitational background satisfying $R_{\mu\nu} = 0$. By
analogy with massless case we restrict ourselves with terms containing
not more than two explicit derivatives (here we treat 
$R_{\mu\nu,\alpha\beta}$ just as external background field). 

We start by investigating all possible two derivative vertexes
containing full four-indexed Riemann tensor and any pair of our four
fields: $\Phi_{\mu\nu\alpha}$, $f_{\mu\nu}$, $A_\mu$ and $\varphi$.

{\bf Vertex} $R D \Phi D \Phi$. Shurely we have to introduce this
vertex which was crucial for the massless case. Here we need only part
containing full Riemann tensor:
\begin{eqnarray}
2m^2 {\cal L}_{12} &=& a_0 R_{\mu\nu,\alpha\beta} [ D^\mu
\Phi^{\nu\rho\sigma} D^\alpha \Phi^{\beta\rho\sigma} +
2 D^\mu \Phi^{\alpha\rho\sigma} D^\rho \Phi^{\sigma\nu\beta} - 
 D^\rho \Phi^{\sigma\mu\alpha} D^\sigma
\Phi^{\rho\nu\beta} - D^\nu \Phi^{\mu\alpha\rho}
D^\beta \tilde{\Phi}^\rho \nonumber \\
 && \qquad -  D^\nu \Phi^{\mu\beta\rho} D^\rho
\tilde{\Phi}^\alpha - D^\mu \tilde{\Phi}^\nu D^\alpha
\tilde{\Phi}^\beta + 
 (D \Phi)^{\mu\alpha} (D \Phi)^{\nu\beta} - 3
(D \Phi)^{\mu\alpha} D^\nu \tilde{\Phi}^\beta ]
\end{eqnarray}
and $\Phi$ field gauge transformations:
\begin{equation}
m^2 \delta_{11} \Phi_{\mu\nu\alpha} = a_0 [ R_{\mu\beta\nu\rho}
D_{[\alpha} \eta_{\beta]\rho} - \frac{2}{d} g_{\mu\nu}
R_{\alpha\beta\rho\sigma} D^\sigma \eta^{\beta\rho} +
(\mu\nu\alpha) ]
\end{equation}

{\bf Vertex} $R \Phi D^2 A$. Complete investigation of this four
derivative 3-2-1 vertex will be given elsewhere, while part
containing full Riemann tensor only is relatively easy to construct.
It has a form:
\begin{equation}
m^2 \Delta_1 {\cal L}_{12} = \frac{6 b_0 (d+1)}{d}
R_{\mu\nu,\alpha\beta}  [ 2 \Phi^{\mu\alpha\rho} D^\nu D^\beta A^\rho
 - 2 \Phi^{\mu\alpha\rho} D^\rho D^\nu A^\beta
 - D^\mu D^\alpha A^\nu \tilde{\Phi}^\beta ]
\end{equation}
and trivially invariant under $A_\mu$ field gauge transformations,
while invariance under $\Phi$ gauge transformations requires that
\begin{equation}
m^2 \delta_{11} A_\mu = b_0 R_{\mu\nu,\alpha\beta} D^\alpha
\xi^{\nu\beta}
\end{equation}

At last, there are tree more cubic vertexes which are trivially gauge
invariant:
\begin{equation}
m^2 \Delta_2 {\cal L}_{12} = R_{\mu\nu,\alpha\beta} [
c_0 F^{\mu\nu,\alpha\beta} \varphi + d_0 ( 4 F^{\mu\nu,\alpha\rho}
f^{\beta\rho} - 4 F^{\mu\alpha} f^{\nu\beta} - 
F^{\mu\nu,\alpha\beta} f) + e_0 D^\mu A^\nu D^\alpha A^\beta ]
\end{equation}
Here $F_{\mu\nu,\alpha\beta}$ --- linearized "curvature" tensor for
$f_{\mu\nu}$ field. The first vertex (with coefficient $c_0$) is
analoge of Gauss-Bonnet vertex multiplied by scalar field but with two
different spin 2 fields. The second one (with coefficient $d_0$) also
comes as the first non-trivial term from such double spin-2
Gauss-Bonnet vertex. Note, that in $d=4$ this vertex is a total
divergency and as a result, as we will see at the end of this Section,
this $d=4$ case is indeed special. The last vertex (with coefficient
$e_0$) is rather well known one containing two gauge invariant field
strengths for vector field $A_\mu$.

Thus we have achieved cancellation of all variations with three
derivatives:
$$
\delta_{01} {\cal L}_{12} + \delta_{11} {\cal L}_{02} = 0
$$
But when $m \ne 0$ we get variations with two derivatives coming from:
$$
\delta_{00} {\cal L}_{12} + \delta_{11} {\cal L}_{01} \ne 0
$$
So we proceed introducing all possible cubic terms with one derivative
into Lagrangian:
\begin{equation}
m {\cal L}_{11} = R_{\mu\nu,\alpha\beta} [ a_1 \Phi^{\mu\alpha\rho}
D_\rho f^{\nu\beta} + a_2 \tilde{\Phi}^\beta D^\nu f^{\mu\alpha} + a_3
\Phi^{\mu\alpha\rho} D^\nu f^{\beta\rho} + a_4 (D \Phi)^{\mu\alpha}
f^{\nu\beta} + a_5 f^{\mu\alpha} D^\nu A^\beta ]
\end{equation}
as well as the only possible correction to gauge transformations:
\begin{equation}
m \delta_{10} f_{\mu\nu} = c_1 R_{\mu\alpha,\nu\beta}
\xi^{\alpha\beta}
\end{equation}
Then requiring that all variations with two derivatives cancel:
$$
\delta_{00} {\cal L}_{12} + \delta_{11} {\cal L}_{01} + \delta_{01}
{\cal L}_{11} + \delta_{10} {\cal L}_{02} = 0
$$
we obtain a number of relations on the coefficients:
$$
b_0 = \frac{(d-2) a_0 + 8 d d_0}{6(d+1)}, \quad
c_0 = - 6 a_0 - \frac{48 d}{d-2} d_0, \quad
e_0 = \frac{2(d+4)(d-1)}{d^2} a_0 + \frac{32(d+2)}{d} d_0
$$
$$
a_1 = a_0 + 16 d_0, \quad a_2 = - 3 a_0 - 8 d_0, \quad
a_3 = - a_0 - 16 d_0, \quad a_4 = - \frac{3}{2} a_0
$$
$$
a_5 = 6 a_0 +  \frac{16(5d+4)}{d} d_0, \quad
c_1 = - \frac{a_0}{2} - 8 d_0
$$
At last, to achieve cancellation of remaining variations with one
derivative and without derivatives we add last correction to
Lagrangian:
\begin{equation}
{\cal L}_{10} = R_{\mu\nu,\alpha\beta} ( b_1 \Phi^{\mu\alpha\rho}
\Phi^{\nu\beta\rho} + b_2 f^{\mu\alpha} f^{\nu\beta} )
\end{equation}
And indeed all variations cancel provided:
$$
b_1 = 2, \qquad b_2 = \frac{9}{2}, \qquad
a_0 = - 4, \qquad d_0 = \frac{1}{4}
$$

Collecting all pieces together we obtain final cubic vertex:
\begin{eqnarray}
m^2 {\cal L}_1 &=& - 4 R_{\mu\nu,\alpha\beta} [ D^\mu
\Phi^{\nu\rho\sigma} D^\alpha \Phi^{\beta\rho\sigma} + 2 D^\mu
\Phi^{\alpha\rho\sigma} D^\rho \Phi^{\sigma\nu\beta} -  D^\rho
\Phi^{\sigma\mu\alpha} D^\sigma \Phi^{\rho\nu\beta} - D^\nu
\Phi^{\mu\alpha\rho} D^\beta \tilde{\Phi}^\rho - \nonumber \\
 && -  D^\nu \Phi^{\mu\beta\rho} D^\rho \tilde{\Phi}^\alpha - D^\mu
\tilde{\Phi}^\nu D^\alpha \tilde{\Phi}^\beta +  (D \Phi)^{\mu\alpha}
(D \Phi)^{\nu\beta} - 3 (D \Phi)^{\mu\alpha} D^\nu \tilde{\Phi}^\beta
] - \nonumber \\
 && - \frac{2 (d-4)}{d} R_{\mu\nu,\alpha\beta} [ 2
\Phi^{\mu\alpha\rho} D^\nu D^\beta A^\rho  - 2 \Phi^{\mu\alpha\rho}
D^\rho D^\nu A^\beta  - D^\mu D^\alpha A^\nu \tilde{\Phi}^\beta ] +
\nonumber \\
 && + 4(d-4) R_{\mu\nu,\alpha\beta} [ \frac{3}{(d-2)} 
F^{\mu\nu,\alpha\beta} \varphi -  \frac{2}{d^2} D^\mu A^\nu D^\alpha
A^\beta ] + \nonumber \\
 && + R_{\mu\nu,\alpha\beta}  [ F^{\mu\nu,\alpha\rho} f^{\beta\rho} -
F^{\mu\alpha} f^{\nu\beta} - \frac{1}{4} F^{\mu\nu,\alpha\beta} f ] +
\nonumber \\
 && + 2m R_{\mu\nu,\alpha\beta} [  5 \tilde{\Phi}^\beta
D^\nu f^{\mu\alpha}  + 3 (D \Phi)^{\mu\alpha} f^{\nu\beta} -
\frac{2(d-4)}{d} f^{\mu\alpha} D^\nu A^\beta ] + \nonumber \\
 && + m^2 R_{\mu\nu,\alpha\beta} ( 2 \Phi^{\mu\alpha\rho}
\Phi^{\nu\beta\rho} + \frac{9}{2} f^{\mu\alpha} f^{\nu\beta} )
\end{eqnarray}
Note, that for $d=4$ this result turns out to be much simpler:
\begin{eqnarray}
m^2 {\cal L}_1 &=& - 4 R_{\mu\nu,\alpha\beta} [ D^\mu
\Phi^{\nu\rho\sigma} D^\alpha \Phi^{\beta\rho\sigma} +
2 D^\mu \Phi^{\alpha\rho\sigma} D^\rho \Phi^{\sigma\nu\beta} - 
 D^\rho \Phi^{\sigma\mu\alpha} D^\sigma
\Phi^{\rho\nu\beta} - D^\nu \Phi^{\mu\alpha\rho}
D^\beta \tilde{\Phi}^\rho - \nonumber \\
 && - D^\nu \Phi^{\mu\beta\rho} D^\rho
\tilde{\Phi}^\alpha - D^\mu \tilde{\Phi}^\nu D^\alpha
\tilde{\Phi}^\beta + 
 (D \Phi)^{\mu\alpha} (D \Phi)^{\nu\beta} - 3
(D \Phi)^{\mu\alpha} D^\nu \tilde{\Phi}^\beta ] + \nonumber \\
 && + 2m R_{\mu\nu,\alpha\beta} [ 
 5 \tilde{\Phi}^\beta D^\nu f^{\mu\alpha} 
+ 3 (D \Phi)^{\mu\alpha} f^{\nu\beta}  ] 
+ m^2 R_{\mu\nu,\alpha\beta} ( 2 \Phi^{\mu\alpha\rho}
\Phi^{\nu\beta\rho} + \frac{9}{2} f^{\mu\alpha} f^{\nu\beta} )
\end{eqnarray}

Thus it is indeed possible (at least at linear approximation) to
extend gauge invariant description of massive spin 3 particles from
constant curvature spaces (such as flat Minkowski space or $(A)dS$) to
arbitrary gravitational background satisfying $R_{\mu\nu} = 0$. It is
instructive to compare our results with those obtained from the
requirement of tree level unitarity \cite{CPD94} (see also recent
paper \cite{GS07}). Due to large  ambiguity one faces working with 
higher derivative interactions it is not an easy task to make direct 
comparison of different formulations (compare e.g. our Lagrangian
(\ref{Lag}) with the Lagrangian (4.28) in \cite{BL06}). Note only,
that the four derivative corrections given in \cite{CPD94} become zero
then one put a "gauge"  $(\partial \Phi)_{\mu\nu} = 0$ and 
$\tilde{\Phi}_\mu = 0$, while it is not to be the case here. Note also
that these corrections have universal form for all spins $s \ge 3$, 
while  general construction for massless particles in $(A)dS$ 
\cite{FV87,FV87a} requires that the number of derivatives grow 
linearly with spin. And indeed, it is not possible to construct gauge
invariant cubic $s-s-2$ vertex for spin $s > 3$ with just four 
derivatives, so the vertex $3-3-2$ constructed in \cite{BL06} and used
here is specific namely for $s=3$.

\vskip 1cm

{\bf Note added.} After the first version of this paper has been 
submitted into arXiv, there appeared the paper \cite{BLS08} where the 
same 3-2-2 vertex was reconstructed in terms of Weyl tensor (again 
due to usual ambiguities their Lagrangian (14) differs  from our 
Lagrangian (\ref{Lag})). Also an appropriate vertex for spin $s=4$ has
been constructed.

\vskip 1cm

{\bf Acknowledgment} \\
Author is gratefull to M. A. Vasiliev for stimulating discussions and
correspondence.

\end{document}